\documentstyle [twocolumn,prb,aps] {revtex}
\begin{document}
\draft
\begin{title}
{Energetics, forces, and quantized conductance in jellium-modeled 
metallic nanowires}
\end{title} 
\author{Constantine Yannouleas, Eduard N. Bogachek, and Uzi Landman }
\address{
School of Physics, Georgia Institute of Technology,
Atlanta, Georgia 30332-0430 }
\date{Physical Review B {\bf 57}, 4872 [1998]}
\maketitle
\begin{abstract}
Energetics and quantized conductance in jellium-modeled nanowires are
investigated using the local-density-functional-based shell correction
method, extending our previous study of uniform-in-shape wires
[C. Yannouleas and U. Landman, J. Phys. Chem. B {\bf 101}, 5780 (1997)]
to wires containing a variable-shaped constricted region. The energetics
of the wire (sodium) as a function of the length of the volume-conserving,
adiabatically shaped constriction, or equivalently its minimum width,
leads to formation of self-selecting magic wire configurations,
i.e., a discrete configurational sequence of enhanced stability, originating
from quantization of the electronic spectrum, namely, formation of
transverse subbands due to the reduced lateral dimensions of the wire.
These subbands are the analogs of shells in finite-size, zero-dimensional
fermionic systems, such as metal clusters, atomic nuclei, and $^3$He
clusters, where magic numbers are known to occur. These variations in the
energy result in oscillations in the force required to elongate the wire
and are directly correlated with the stepwise variations of the conductance
of the nanowire in units of $2e^2/h$. The oscillatory patterns in the
energetics and forces, and the correlated stepwise variation in the 
conductance are shown, numerically and through a semiclassical analysis,
to be dominated by the quantized spectrum of the transverse states at
the narrowmost part of the constriction in the wire.
\end{abstract}
~~~~\\

\narrowtext

\section{Introduction}

Understanding of the physical origins and systematics
underlying the variations of materials properties with size, form of 
aggregation, and dimensionality are some of the main challenges in modern
materials research, of ever increasing importance  in the face of the
accelerated trend toward miniaturization of electronic and mechanical
devices. \cite{mart1,issp,avou,nano} 

Interestingly, it has emerged that concepts and methodologies developed in the
context of isolated gas-phase clusters and atomic nuclei are often most useful
for investigations of finite-size solid-state structures. In particular, it has
been shown most recently 
\cite{land4,barn} 
through first-principles molecular dynamics
simulations that as metallic (sodium) nanowires are stretched to just a few
atoms in diameter, the reduced dimensions, increased surface-to-volume ratio, 
and impoverished atomic environment, lead to formation of structures, made of
the metal atoms in the neck, which can be described in terms of those 
observed in small gas-phase sodium clusters; hence they were termed 
\cite{land4,barn} 
as supported {\it cluster-derived structures (cds)\/}. The above prediction of
the occurrence of ``magic-number'' cds's in nanowires, due to characteristics
of electronic cohesion and atomic bonding in such structures of reduced
dimensions, are directly correlated with the energetics of metal clusters,
where magic-number sequences of cluster sizes, shapes and structural motifs 
due to electronic and/or geometric shell effects, have been long predicted and 
observed. \cite{heer,yann1,mart}

These results lead one directly to conclude that other properties
of nanowires, derived from their energetics, may be described using 
methodologies developed previously in the context of clusters. Indeed, in a
previous letter, \cite{yann5} we showed that certain aspects of the mechanical 
response (i.e.,
elongation force) and electronic transport (e.g., quantized conductance) in
metallic nanowires can be analyzed using the local-density-approximation
(LDA) -based shell correction method (SCM), developed and applied previously
in studies of metal clusters. \cite{yann1,yann2}
Specifically, we showed that in a jellium-modelled, volume-conserving, and
uniform in shape nanowire, variations of the total energy (particularly 
terms associated with electronic subband corrections) upon elongation of the
wire lead to {\it self-selection\/} of a sequence of stable ``magic''
wire configurations (MWC's, specified by a sequence of the wire's
radii), with the force required to elongate the wire from one configuration to 
the next exhibiting an oscillatory behavior. 
Moreover, we showed that due to the
quantized nature of electronic states in such wires, the electronic conductance
varies in a quantized step-wise manner (in units of the conductance quantum
$g_0=2e^2/h$), correlated with the transitions between MWC's and the 
above-mentioned force oscillations.

In this paper, we expand our LDA-based treatment to wires of variable
shape, that is allowing for a constricted region. From this investigation,
we conclude that the above self-selection principles and the direct 
correlations between the oscillatory patterns in the energetic stability,
forces, and stepwise variations of the quantized conductance maintain for
the variable-shaped wire as well, with the finding that underlying these
oscillatory patterns and 
correlations are the contributions from the narrowmost region of the wire.
Furthermore, this finding is analyzed and corroborated through a semiclassical
analysis.

Prior to introducing the model studied in this paper, it is appropriate to 
briefly describe certain previous theoretical and experimental investigations,
which form the background and motivation for this study. Atomistic
descriptions, based on realistic interatomic interactions, and/or 
first-principles modelling and simulations played an essential role in
discovering the formation of nanowires, \cite{land1} and in 
predicting and elucidating
the microscopic mechanisms underlying their mechanical, spectral, electronic
and transport properties. 

These predictions \cite{land1,land2,land3} [particularly those
pertaining to generation of nanowires through separation of the contact
between two materials bodies; size-dependent evolution of the wire's
mechanical response to elongation transforming from multiple-slips for
wider wires to a succession of stress accumulation and fast relief stages 
leading to a sequence of structural instabilities and order-disorder
transformations localized in the neck region when its diameter shrinks
to about 15 \AA; consequent oscillations of the elongation force and the
calculated high value of the resolved yield stress ( $\sim$ 4 GPa for
Au nanowires; which is over an order of magnitude that of the bulk), as well 
as anticipated electronic quantization effects on transport
properties \cite{land1,boga1}]
have been corroborated in a number of experiments using 
scanning tunneling and force microscopy, 
\cite{land1,pasc1,oles,pasc2,smith,rubi,stal} 
break junctions, \cite{krans} 
and pin-plate techniques \cite{land2,costa}
at ambient environments, as well as under 
ultrahigh vacuum and/or cryogenic conditions. Particularly pertinent
to our current study are experimental observations of the oscillatory
behavior of the elongation forces and the correlations between the changes in 
the conductance and the force oscillations; see especially the simultaneous
measurements of force and conductance in gold nanowires in Ref.\ 
\onlinecite{rubi},
where in addition the predicted ``ideal'' value of the critical yield
stress has also been measured (see also Ref.\ \onlinecite{stal}).

The LDA-jellium-based model introduced in our previous paper \cite{yann5}
and extended to generalized wire shapes herein, 
while providing an appropriate solution within the model's assumptions
(see section II),
is devoid by construction of atomic crystallographic
structure and does not address issues pertaining to nanowire formation methods,
atomistic configurations, and mechanical response modes [e.g., 
plastic deformation mechanisms, interplanar slip, ordering and disordering 
mechanisms (see 
detailed descriptions in Refs.\ \onlinecite{land1,land2}
and \onlinecite{land3}, 
and a discussion of conductance dips in Ref.\ \onlinecite{pasc2}),
defects, mechanical reversibility, \cite{rubi,land2}
and roughening of the wires's morphology 
during elongation \cite{land3}], nor does it consider the effects of the above 
on the electron spectrum, transport properties, and dynamics. \cite{barn}
Nevertheless, as shown below, the model offers a useful framework for linking 
investigations of solid-state 
structures of reduced dimensions (e.g., nanowires) with methodologies
developed in cluster physics, as well as highlighting certain nanowire
phenomena of mesoscopic origins and their analogies to clusters.

In this context, we note that several other treatments related to certain
of the issues in this paper, but employing free-electron models,
have been pursued most recently. \cite{ruit,staf} In both of these 
treatments an infinite confining potential on the surface of the
wire is assumed and only the contribution from the kinetic
energy of the electrons to the total energy is considered, neglecting
the exchange-correlation and Hartree terms, and electrostatic interactions
due to the positive ionic (jellium) background.
A comprehensive discussion of the limitations of such free-electron models
in the context of calculations of electronic structure and energetics 
(e.g., surface energies) of metal surfaces can be found in Ref.\ 
\onlinecite{lang3}.

In section II.A., we outline the LDA-based Shell Correction Method,
describe the jellium model for variable-shaped nanowires, and derive
expressions for the energetics of such nanowires (density of states, energy,
and force). Numerical results pertaining to energetics, force, and electronic
conductance, calculated as a function of elongation 
for variable-shaped sodium nanowires, 
are given in section II.B., including a discussion on the main finding that
the contribution from the narrowmost part of the constriction underlies the
properties of these quantities and the correlations between them.
These correlations between the energetic and transport properties and
their dependence on the narrowmost part of the nanowire are further
analyzed in section III, using a semiclassical treatment. We summarize our
results in section IV. 

\section{Density Functional Description of Jellium Nanowire}

\subsection{Theory}

\subsubsection{Shape of Constriction}
Consider a jellium nanowire with circular symmetry about the axis of the
wire ($z$ axis). The wire may contain a constricted region (see Fig.\ 1), 
that is a section of length $L$ where the cross-sectional radius $a(z)$ varies
along the axis as
\begin{equation}
a(z) = a_0 + (R_0-a_0) f(z)\;,\;\ -L/2 \leq z \leq L/2~,
\label{az}
\end{equation}
with $f(-z)=f(z)$ (the $z=0$ plane passes through the middle of the wire) and
$f(\pm L/2)=1$. $R_0=a(\pm L/2)$ is the uniform radius outside the constricted
section, and $a_0 \equiv a(0)$. In this paper, we take a parabolic shape
$f(z)=(2z/L)^2$ for the description of the constricted region [a wire of
uniform cross section throughout corresponds to $f(z)=1$].

We also assume that elongation of the wire occurs in the constricted region
while maintaining its volume constant (this is supported by MD simulations), 
namely by requiring that 
\begin{equation}
2 \int_0^{L/2} a^2(z) dz= R_0^2 L_0~,
\label{vol}
\end{equation}
for given values of $R_0$ and $L_0$ [hereafter we will denote the pair of
parameters ($R_0$,$L_0$) by ${\cal O}$; we further assume that $R_0 \ll L_0$]. 
For the parabolic shape assumed in this paper,
the smallest cross-sectional radius is determined
for any given value of $L_0 \leq L \leq 5 L_0$ from Eqs.\ (\ref{az}) 
and (\ref{vol}) as
\begin{equation}
a_0 = \frac{R_0}{4} \left[-1 + (30 \frac{L_0}{L} -5)^{1/2} \right]~,
\label{a0}
\end{equation}
i.e., $a_0=R_0$ for $L=L_0$, and $a_0=0$ (i.e., breakage of the wire)
for $L=5 L_0$. 

\subsubsection{Shell Correction Method}

The Shell Correction Method we employ is based on the LDA theory. 
In the Shell Correction Method \cite{yann1,yann2,yann3,yann4} (SCM), 
the total LDA energy, $E_T(L,{\cal O})$, for any configuration of the wire 
(specified by $L$ and ${\cal O}$) is separated as,
\begin{equation}
E_T(L,{\cal O}) =
\widetilde{E}(L, {\cal O}) +
\Delta E_{\text{sh}} (L,{\cal O})~,
\label{etot}
\end{equation}
where 
$\widetilde{E}(L, {\cal O})$
varies smoothly as a function of the system size ($L$) while
$\Delta E_{\text{sh}} (L,{\cal O})$ varies in an oscillatory manner with
$L$, as a result of the quantization of the electronic states.
$\Delta E_{\text{sh}} (L,{\cal O})$ is 
usually called a shell correction in the nuclear \cite{stru,bm}
and cluster \cite{yann1,yann2} literature; we
continue to use here the same terminology with the understanding that the 
electronic levels in the nanowire form subbands, which are the analog of
electronic shells in clusters where the size of the system is 
usually given by specifying the number of atoms $N$. The SCM method,
which has been shown to yield results in excellent agreement with experiments
\cite{yann1,yann3,yann4}
and self-consistent LDA calculations 
\cite{yann1,yann2} for a number of cluster systems, is 
equivalent to a Harris functional
($E_{\text{harris}}$) 
approximation to the Kohn-Sham LDA with the input density,
$\rho^{\text{in}}$, obtained through variational
minimization of an extended Thomas-Fermi (ETF) energy functional,
$E_{\text{ETF}}[\rho]$.

The Harris functional is given by the following expression,
\begin{eqnarray}
E&&_{\text{harris}} [\rho^{\text{in}}] = E_I + \sum_{i=1}^{\text{occ}} 
\epsilon_i^{\text{out}} 
\nonumber \\
&& -  \int \! \left\{ \frac{1}{2} V_H [ \rho^{\text{in}} ({\bf r})] 
+ V_{\text{xc}} [ \rho^{\text{in}} ({\bf r})] \right\} 
\rho^{\text{in}} ({\bf r}) d{\bf r} 
\nonumber \\
&& + \int \! {\cal E}_{\text{xc}} [ \rho^{\text{in}} ({\bf r})] d{\bf r}~,
\label{enhar}
\end{eqnarray}
where $V_H$ is the Hartree (electronic) repulsive potential, 
$E_I$ is the repulsive electrostatic energy of the ions, and
$E_{\text{xc}} \equiv \int {\cal E}_{{xc}} [\rho] d {\bf r}$ 
is the exchange-correlation (xc) functional \cite{gunn}
[the corresponding xc potential is given as $V_{\text{xc}}({\bf r}) \equiv
\delta E_{\text{xc}} [\rho] / \delta \rho({\bf r})$].
$\epsilon_i^{\text{out}}$ are the eigenvalues
(non-self-consistent) of the single-particle Hamiltonian,
\begin{equation}
\widehat{H} = - \frac{\hbar^2}{2m_e} \nabla^2 + V_{\text{in}}~,
\label{hin}
\end{equation}
with the mean-field potential given by
\begin{equation}
V_{\text{in}} 
[\rho^{\text{in}} ({\bf r})] =
V_H [\rho^{\text{in}} ({\bf r})]
+ V_{\text{xc}} [\rho^{\text{in}} ({\bf r})] 
+ V_I ({\bf r})~,
\label{mfpot}
\end{equation}
$V_I ({\bf r})$ being the attractive potential between the electrons and
ions.

In electronic structure calculations where the corpuscular nature of the ions 
is included (i.e., all-electron or pseudo-potential calculations), 
$\rho^{\text{in}}$ may be taken as a superposition of atomic-site densities. 
In the case of jellium calculations, we have 
shown \cite{yann1,yann2} that an accurate approximation to the KS-LDA total 
energy is obtained by using the Harris functional with the input density, 
$\rho^{\text{in}}$, in Eq.\ (\ref{enhar}) evaluated from a variational 
Extended-Thomas-Fermi (ETF)-LDA calculation.

The ETF-LDA energy functional, $E_{\text{ETF}} [\rho]$, 
is obtained by replacing the kinetic energy term,
$T[\rho]$, in the usual LDA functional, namely in the expression,
\begin{eqnarray}
E&&_{\text{LDA}}[\rho]=T[\rho]  \nonumber \\
&& + \int \left\{ \frac{1}{2} V_H [\rho({\bf r})]
  + V_I({\bf r}) \right\} \rho({\bf r})  d\/ {\bf r} \nonumber \\
&& + \int {\cal E}_{\text{xc}} [\rho({\bf r})]d\/ {\bf r} + E_I~,
\label{enlda}
\end{eqnarray}
by the ETF kinetic energy, given to the 4th-order gradients as follows,
\cite{hodg}
\begin{eqnarray}
&& \frac{2m_e}{\hbar^2} 
T_{\text{ETF}}[\rho] =
\frac{2m_e}{\hbar^2} 
\int t_{\text{ETF}}[\rho] d {\bf r} = \nonumber \\
&& = \int \! \left\{
\frac{3}{5} (3\pi^2)^{2/3} \rho^{5/3} + 
\frac{1}{36} \frac{(\nabla \rho)^2}{\rho} +
\frac{1}{270} (3 \pi^2)^{-2/3} \rho^{1/3}  
\right. \nonumber \\
&& \times 
\left. \left[ 
\frac{1}{3} \left( \frac{\nabla \rho} {\rho}
\right)^4 - \frac{9}{8} \left( \frac{\nabla \rho} {\rho} \right)^2 
\frac{\Delta \rho} {\rho} +
 \left( \frac{\Delta \rho} {\rho} \right)^2 
\right] \right\}
d {\bf r}~.
\label{t4th}
\end{eqnarray}

The optimal ETF-LDA total energy is then obtained by minimization of
$E_{\text{ETF}} [\rho]$ 
with respect to the density. In our calculations, we use for the trial
densities parametrized profiles $\rho ({\bf r};\; \{\gamma_i\})$ with
$\{ \gamma_i \}$ as variational parameters (the ETF-LDA optimal density is
denoted as $\widetilde{\rho}$). The single-particle eigenvalues,
$\{\epsilon_i^{out}\}$, in Eq. (\ref{enhar})
are obtained then as the solutions to the single-particle Hamiltonian
of Eq.\ (\ref{hin}) with $V_{\text{in}}$ replaced by $V_{\text{ETF}}$
[given by Eq.\ (\ref{mfpot}) with $\rho^{\text{in}} ({\bf r})$
replaced by $\widetilde{\rho} ({\bf r})$].
Hereafter, these single-particle eigenvalues will be denoted by 
$\{ \widetilde{\epsilon}_i \}$.

In our approach,
the smooth contribution in the separation (\ref{etot}) of the total
energy is given by $E_{\text{ETF}} [\widetilde{\rho}]$, while
the shell correction, $\Delta E_{\text{sh}}$, is simply the difference
\begin{eqnarray}
\Delta E_{\text{sh}} && = E_{\text{harris}}[\widetilde{\rho}]
- E_{\text{ETF}} [\widetilde{\rho}] \nonumber \\
&& = \sum_{i=1}^{\text{occ}} \widetilde{\epsilon}_i -
\int \! \widetilde{\rho}({\bf r}) V_{\text{ETF}} ({\bf r}) d\/ {\bf r}
- T_{\text{ETF}} [\widetilde{\rho}]~.
\label{dsh}
\end{eqnarray}

\subsubsection{Adiabatic Assumption}

The volume density of the positive background is given by
$\rho^+_v = 3/(4\pi r_s^3)$, where $r_s$ is the Wigner-Seitz radius
characteristic to the material, and thus the number of positive charges
in the constriction is
\begin{equation}
N^+ ({\cal O})=3R_0^2 L_0/(4 r_s^3)~.
\label{numpos}
\end{equation}

Since the nanowire contains a constricted region of variable 
cross-sectional radius $a(z)$ [see Eq.\ (\ref{az})], we define a linear
(i.e., density per unit length of the nanowire) background density
$\rho^+_l (z;L,{\cal O}) = 3 a^2 (z)/(4 r_s^3)$, 
which when integrated over the total length of the wire yields
$N^+ ({\cal O})$ [see Eq.\ (\ref{numpos})]. Correspondingly, the
variational electronic volume density 
$\widetilde{\rho} ({\bf x}; L, {\cal O}) \equiv
\widetilde{\rho} (r, z ; L, {\cal O})$, and in our calculations it takes 
the form,
\begin{equation}
\widetilde{\rho} (r,z; L, {\cal O}) = 
\frac{\widetilde{\rho}_0(z)}{ \left[ 1+
\exp \left(\frac{r-r_0(z)}{\alpha(z)}\right) \right]^{\gamma(z)}}~,
\label{rho}
\end{equation}
with $\widetilde{\rho}_0 (z)$, 
$\alpha(z)$, and $\gamma(z)$ as $z$-dependent 
variational parameters. In the ETF calculation, 
$\widetilde{\rho}$ is determined variationally at a given $z$ as the one
associated with a uniform cylinder of radius $a(z)$ (adiabatic assumption),
under the normalization condition for local charge neutrality, namely,
$2 \pi \int dr [r \widetilde{\rho} (r,z; L, {\cal O})] = 
\rho^+_l (z; L, {\cal O})$ [which fixes the 4th parameter
$r_0 (z)$ in Eq.\ (\ref{rho})]. 
The optimized
$\widetilde{\rho}$ 
allows then calculation of the smooth contribution for any length of the
constriction,
$\widetilde{E} (L, {\cal O}) \equiv 
E_{\text{ETF}} (L, {\cal O})$ 
in Eq.\ (\ref{etot}).

The calculation of the shell-correction term,
$\Delta E_{\text{sh}} (L, {\cal O})$, in Eq.\ (\ref{etot})
proceeds by evaluating first the density of states in the nanowire.
Assuming an adiabatic separation of the ``fast'' transverse and the ``slow'' 
longitudinal variables, \cite{boga1,glaz,imry} the electronic wave functions 
in the classically allowed regions may be written as
\begin{eqnarray}
\Psi_{nm\epsilon} (r,\phi,z; && L, {\cal O}) \propto
\psi_{nm} (r; z, L, {\cal O}) e^{i m \phi} \nonumber \\
&& \times 
e^{i \int^z dz^\prime k^{nm}_\perp (z^\prime ; \epsilon, L, {\cal O})}~,
\label{wvf}
\end{eqnarray}
where
$k_\perp^{nm}$ is the local wave number along the axial ($z$) direction of
the nanowire
\begin{equation}
k^{nm}_\perp (z, \epsilon; L, {\cal O}) =
\left[ \frac{2m_e}{\hbar^2}
[ \epsilon - 
\widetilde{\epsilon}_{nm} (z; L, {\cal O}) ] \right]^{1/2}~,
\label{kperp}
\end{equation} 
and $\widetilde{\epsilon}_{nm}$ is the (transverse) 
local eigenvalue spectrum at
$z$. To calculate this spectrum for a wire of a configuration specified
by $(L, {\cal O})$ for any value of $z$, the eigenvalues of a cylindrical
wire with a (uniform) radius $a(z)$ are calculated from the two-dimensional
Schr\"{o}dinger equation
\begin{eqnarray}
-\frac{\hbar^2}{2m_e} \left[ \frac{d^2~}{dr^2} \right.
&& \left. +\frac{1}{r} \frac{d~}{dr} - \frac{m^2}{r^2} \right] \psi +
V_{\text{ETF}}(r; z,L, {\cal O}) \psi \nonumber \\
&& = \widetilde{\epsilon}_{nm} (z, L, {\cal O}) \psi~.
\label{scheq}
\end{eqnarray}

The linear (per unit length), one-dimensional density of states at $z$,
$D_l (z, \epsilon; L, {\cal O})$, 
is given by
\begin{eqnarray}
D_l (z, \epsilon; L, {\cal O}) && =
\frac{2}{\pi} \sum_{nm} 
\frac{\partial k_{\perp}^{nm} (z, \epsilon; L, {\cal O})}
{\partial \epsilon} \nonumber \\ 
&& \times \Theta[ \epsilon - \widetilde{\epsilon}_{nm}(z; L, {\cal O})]~,
\label{elden}
\end{eqnarray}
where spin degeneracy has been included, and $\Theta$ is the
Heaviside step function.

From Eq.\ (\ref{kperp}), we obtain
\begin{eqnarray}
D_l (z, \epsilon; L, {\cal O}) && =
 \left( \frac{2m_e}{\pi^2 \hbar^2} \right)^{1/2} 
 \sum_{nm} \left[  \epsilon - 
\widetilde{\epsilon}_{nm} (z; L, {\cal O}) \right]^{-1/2} \nonumber \\
&& \times \Theta[ \epsilon - \widetilde{\epsilon}_{nm}(z; L, {\cal O}) ]~.
\label{elden2}
\end{eqnarray}

We may now define an integrated density of states in the constriction
\begin{equation}
D (\epsilon; L, {\cal O}) =
\int_{-L/2}^{L/2} dz D_l (z, \epsilon; L, {\cal O})~.
\label{eldt}
\end{equation}

The total number of states up to energy $\epsilon$ 
in the constricted region of the wire is given by
\begin{eqnarray}
N&&^{-} (\epsilon; L, {\cal O}) = 
\int_0^\epsilon d\epsilon^\prime D (\epsilon^\prime; L, {\cal O}) 
\nonumber \\
&& = \frac{2}{\pi} \int_{-L/2}^{L/2} dz 
\sum_{nm} 
\sqrt{ \frac{2m_e}{\hbar^2} [\epsilon - \widetilde{\epsilon}_{nm}
(z; L, {\cal O}) ] } \nonumber \\
&& \times \Theta[ \epsilon - \widetilde{\epsilon}_{nm}(z; L, {\cal O}) ]~.
\label{totnum}
\end{eqnarray}

Since the total number of electrons in the constricted region is
$N^+ ({\cal O})$ [see Eq.\ (\ref{numpos})], the Fermi energy,
$\epsilon_F (L, {\cal O})$, for a wire with a configuration specified by
($L, {\cal O}$) is given from Eq.\ (\ref{totnum}), i.e., 
\begin{equation}
N^{-}(\epsilon_F; L, {\cal O}) =
N^+ ({\cal O})~.
\label{fermi}
\end{equation}

Using the above and Eq.\ (\ref{dsh}), the shell-correction term, 
\begin{equation}
\Delta E_{\text{sh}} (L, {\cal O}) 
\equiv
E_{\text{harris}} [\widetilde{\rho}; L, {\cal O}] -
E_{\text{ETF}} [\widetilde{\rho}; L, {\cal O}]~,
\label{harr}
\end{equation}
may be calculated as
\begin{eqnarray}
\Delta && E_{\text{sh}} (L, {\cal O}) =
\int_0^{\epsilon_F (L, {\cal O})}
d\epsilon [\epsilon D(\epsilon; L, {\cal O})] \nonumber \\
 && -2 \pi \int_{-L/2}^{L/2} dz \int_0^\infty dr
r \widetilde{\rho} (r,z; L, {\cal O})
V_{\text{ETF}} (r,z; L, {\cal O}) \nonumber \\
 && -2 \pi \int_{-L/2}^{L/2} dz \int_0^\infty dr
r t_{\text{ETF}} [\widetilde{\rho} (r,z; L, {\cal O})]~,
\label{dshw}
\end{eqnarray}
where $V_{\text{ETF}}$ is the ETF potential (Hartree, exchange-correlation,
and electron attraction to the positive background) and
$t_{\text{ETF}}$ is the volume density of the ETF kinetic-energy functional
[see Eq.\ (\ref{t4th})].

In actual calculations, we invert the order of integration in the
first term of Eq.\ (\ref{dshw}), which then takes the form
\begin{eqnarray}
&&\frac{2}{3\pi} 
\int_{-L/2}^{L/2} dz
\sum_{nm}
[ \epsilon_F + 2 \widetilde{\epsilon}_{nm} (z; L, {\cal O}) ] 
\nonumber \\
&& \times \sqrt{ \frac{2m_e}{\hbar^2} [\epsilon_F - \widetilde{\epsilon}_{nm}
(z; L, {\cal O}) ] } \;
\Theta[ \epsilon_F - \widetilde{\epsilon}_{nm}(z; L, {\cal O}) ]~.
\label{sumi}
\end{eqnarray}

Note that Eq.\ (\ref{fermi}) implies a common Fermi level for the whole
constriction for a given $L$ (i.e., $\epsilon_F$ is not a local property).
Therefore, Eq.\ (\ref{sumi}) is not equivalent to integration of the 
corresponding uniform wire result derived by us in Ref.\ 
\onlinecite{yann5} over the $z$-coordinate, since there $\epsilon_F$
varies with the wire's cross-sectional radius.

Having calculated the smooth and shell-correction contributions to the 
total energy, as a function of $L$, the total elongation force may be evaluated
as the derivative of the total energy with respect to $L$, i.e.,
$F_T = - dE_T/dL$, and the contributions to it from the smooth and
shell-correction terms are given by 
$\widetilde{F} = - d{\widetilde{E}}/dL$, and
$\Delta F_{\text{sh}}= - d \Delta E_{\text{sh}}/dL$.

\subsection{Results}

In this section, we report results for the elongation of a Sodium nanowire
($r_s=4$ a.u.),
starting with an initial cylindrical constriction of length $L_0=80$ a.u.
and radius $R_0=25$ a.u. 

In Fig.\ 2 , we show electronic-potential profiles,
$V_{\text{ETF}}[r; a(z), L, {\cal O}]$, 
for a particular constriction with $\Delta L /L_0 =1.125$ ($\Delta L=L-L_0$).
We display here the potential profiles calculated at the narrowmost 
part of the constriction [$a_0 \equiv a(0) = 12.62$ a.u.] and at
its end [i.e., at $a(L/2)=R_0$]
(in this paper, all the subsequent numerical results we will discuss relate to 
constrictions with the same set of ${\cal O}$-parameters, namely, 
$L_0=80$ a.u. and $R_0=25$ a.u.). 
We found that for other values of $z$ (i.e., for $ 0 < |z| < L/2$), 
the potential assumes profiles intermediate between the two profiles shown
here, namely, the depth of the potential well remains practically
unaltered, while its width follows the enlargement of the
jellium-background radius $a(z)$, from $a_0$ to $a(L/2)$.
From the three components, $V_H$, $V_I$, and $V_{\text{xc}}$, which
contribute to the total $V_{\text{ETF}}$ [see Eq.\ (\ref{mfpot})],
we found that for all these potential profiles, calculated for different
values of $z$ along the constriction,
the xc contribution is the dominant one, amounting to
approx. $-5.4$ eV, while the total electrostatic
contribution, $V_H + V_I$, is much smaller, resulting in a characteristic
``winebottle'' profile familiar from LDA studies of spherical clusters.
\cite{wine} 

Fig.\ 2 also displays the transverse local eigenvalues 
$\widetilde{\epsilon}_{nm}$ 
associated with the two potential profiles. Naturally, a wider 
potential profile yields a larger number of such eigenvalues below the Fermi 
level. 


To illustrate the nature of the electronic spectrum in the nanowires, and its
dependencies on the characteristics of the wire, i.e., shape and length,
we show in Fig.\ 3(a) densities of states, $D (\epsilon; L, {\cal O})$, 
calculated for a variable-shaped wire
for two wire lengths (and consequently two minimal 
cross-sectional radii of the constricted region). 
The density of states for a uniform wire with a radius equal to that
of the unconstricted region of the variable-shaped ones [shown in Fig.\ 
3(a)] is displayed in Fig.\ 3(b). 

Two ``classes'' of features are noted
for the variable-shaped wires: (i) those associated with the narrowmost
constricted region (marked by numbers) whose radius, $a_0$, varies upon
elongation, and (ii) those associated with the maximal radius of the
constriction (and with the unconstricted part of the wire) which remains
constant throughout the elongation of the wire. Identification of the latter
class of features (several of which are marked by arrows) is facilitated
through comparison with the density of states for the corresponding
uniform wire [Fig.\ 3(b)]. We observe here that, for the broader
(and thus shorter) wire [lower curve in Fig.\ 3(a)], six of the
features (peaks) in the density of states coming from the spectrum of
transverse energy levels at the narrowmost region of the constriction
are located below the Fermi level, $\epsilon_F$ [all the peaks in the
density of states occur at the energies of the transverse levels; e.g., 
compare the location of the peaks in the
lower curve in Fig.\ 3(a) with the corresponding spectrum
on the left side of Fig.\ 2].

On the other hand, for the much narrower (and thus longer) constricted
wire, only one of these peaks is below $\epsilon_F$ [see upper curve in
Fig.\ 3(a)]. When plotting the density of states at $\epsilon_F$ versus
the elongation (or equivalently the minimal radius of the constricted
region), these variations lead to an oscillatory pattern, as peaks in the
density of states are shifted above the Fermi level, one after the other
as the wire is being elongated. These variations are also portrayed
in the energetics of the wire (shown in Fig.\ 4), and 
in the stepwise behavior of the quantized conductance through the wire
versus length (see Fig.\ 5 below).

From Fig.\ 4, we observe that the magnitude of the smooth ETF contribution,
$\widetilde{E}$, to the total energy, $E_T$, of the wire is dominant, 
with the shell-correction contribution, $\Delta E_{\text{sh}}$,
exhibiting an oscillatory pattern, with local minima at a set
of wire lengths (and correspondingly a set of minimal cross-sectional 
radii) which we term ``magic wire configurations'' (MWC's), i.e., wire 
configurations with enhanced energetic stability. When added to the smooth
contribution, these shell-correction features lead to local minima of the
total energy toward the end of the elongation (and consequently, narrowing)
process, while for thicker wires (i.e., $\Delta L/L_0 \leq 2.5$ in 
Fig.\ 4) they are expressed as inflection points of the total energy 
(in this context, see the total-force curve, $F_T$, in Fig.\ 5, where
the local minimum in $E_T$ corresponds to the point with $F_T=0$ marked
by an arrow).

We note here that the occurrence of local minima in the total energy 
results from a balance between $\Delta E_{\text{sh}}$ and
$\widetilde{E}$, with the latter increasing (that is acquiring less
negative values) as the constriction elongates due to the increasing
contribution from the surface of the constriction.
Comparison of the magnitudes of the
shell corrections in a variable-shaped wire and in a uniform one [i.e.,
one with $f(z) \equiv 1$ in Eq.\ (\ref{az}), whose case was discussed 
in Ref.\ \onlinecite{yann5}] shows that the amplitudes of the oscillations
in the latter case are much larger (over an order of magnitude). The
reason for this difference is that in the constant-radius 
wire the quantization 
into the transverse subbands is uniform along the wire, while in the
variable-shaped case the subband spectrum is different in various parts of
the constriction. While the oscillatory pattern is dominated by the
spectrum at the narrowmost region (see also Section III below), the
amplitudes are influenced by the transverse-mode spectra from other parts
of the constriction. Consequently, the number of local minima in the
total energy, $E_T$, (and thus the number of wire configurations, i.e.,
lengths, for which the total force, $F_T$, vanishes) is larger for a
uniform wire than for a variable-shaped one. Additionally, we suggest that
for materials with relatively smaller surface energies a larger number 
of local minima may occur.

From the total energy, and the smooth and shell-correction contributions to
it, we obtain the total ``elongation force'' (EF), $F_T$, and the
corresponding components of it, $\widetilde{F}$ and
$\Delta F_{\text{sh}}$. These results are displayed in Fig.\ 5, along with
the conductance of the wire evaluated, in the adiabatic approximation
(i.e., no mode mixing \cite{glaz}) and neglecting tunneling effects
(assuming unit transmission coefficients for all the conducting modes),
using the Landauer expression, \cite{landa,imry2}
\begin{equation}
G(L, {\cal O}) = g_0 \sum_{nm} 
\Theta [\epsilon_F - \widetilde{\epsilon}_{nm}
(z=0; L, {\cal O})]~,
\label{land}
\end{equation}
where $g_0=2 e^2/h$, and the spectrum of the transverse modes is evaluated
(for each constriction length) at the narrowmost part of the constriction,
$z=0$. Tunneling contributions (see e.g., Ref.\ \onlinecite{boga0}), 
mode-mixing and non-adiabaticity may affect the sharpness of the conductance 
steps, and/or introduce some interference related features, particularly
near the transitions between the conductance plateaus. These effects, which
can be included in more elaborate evaluations of the conductance,
\cite{lang,brand,todo} do not modify the conclusions of our study.

Also included in this figure is a plot describing the variation of
the minimal cross-sectional radius $a_0$ with the length of the
constriction [see Eq.\ (\ref{a0})].

As evident from Fig.\ 5, the oscillations in the force resulting from the
shell-correction contributions are prominent. In $\Delta F_{\text{sh}}$, we
observe that the locations of the zeroes of the force situated at the 
right of the force maxima occur for values of $\Delta L /L_0$ which
coincide with the locations of local minima in the shell-correction 
contribution to the energy of the wire (i.e, for a sequence of minimal 
cross-sectional radii corresponding to MWC's). In the total force,
$F_T$, only one of these points (where $F_T=0$) remains [i.e., the one
corresponding to the local minimum in the total energy towards the end
of the elongation process (see Fig.\ 4)], for the reasons discussed 
above in connection with the energetics of the wire. Nevertheless, the
oscillations in the total force correlate well with those in the total
energy of the wire, which as discussed above originate from the 
subband spectrum at the narrowmost part of the constriction 
(see also section III). Also, the locations of the local maxima in the
total force correlate with the stepwise variations in the conductance 
signifying the sequential decrease in the number of transverse 
subbands (calculated at the narrowmost section of the wire) below
$\epsilon_F$ (i.e., conducting channels) as the constricted part of
the wire elongates (and thus narrows). Additionally, we note that the
magnitude of the total force is comparable to measured ones (i.e., in
the nanonewton range). The magnitude of the total force in sodium
nanowires (not measured to date) is expected to be smaller than that
found for gold nanowires, \cite{rubi,stal} due mainly to 
differences in the electron densities and surface energies of the materials.

\section{Semiclassical Analysis}

As discussed above, the total energy of the wire is characterized by local
minima and inflection points occuring for a set of wire lengths, or
equivalently a set of minimal cross-sectional radii of the constriction,
and are reflected in the oscillatory patterns of the elongation force.
These features correspond to the oscillatory shell-correction contributions
and originate from the spectrum of transverse modes at the narrowmost
part of the constriction. Moreover, these patterns correlate
with the locations of the quantized conductance steps, which are determined
by the transverse-mode spectrum at the narrowmost region (i.e., the
number of conducting modes below $\epsilon_F$, and their degeneracies).

To further investigate the origins of these correlations, 
we present in this section
a semiclassical analysis of the density of states, energetics, forces,
and conductance in a
free-electron nanowire modeled via an infinite confining potential on the 
surface of the wire. As in the above (see Fig.\ 1), we model the constricted
region of the wire as a section with a slowly (adiabatically) varying
shape. Dividing the constriction into thin cylindrical slices, the solution
of the Schr\"{o}dinger equation for each slice is of the form,
\begin{equation}
\psi = {\cal A} J_m (\kappa r) e^{im \phi} e^{ip_\perp z}~,
\label{bess}
\end{equation}
where ${\cal A}$ is a normalization constant, $p_\perp$ is the electron
momentum along the axis of the wire, $J_m (\kappa r)$ is the Bessel
function of order $m$, and 
$\kappa = (2m_e \epsilon -p_\perp^2)^{1/2}/\hbar$.

Consider first a uniform cylindrical wire with a constant cross-sectional
radius $a$. With the infinite wall boundary condition assumed here, the 
single-particle electronic energy levels in the wire are expressed in terms
of the roots of the Bessel functions, $\gamma_{nm}$, as
\begin{equation}
\epsilon_{nm,p_\perp} =
\frac{\hbar^2 \gamma_{nm}^2}{2 m_e a^2} +
\frac{p_\perp^2}{2m_e}~.
\label{ebes}
\end{equation}

Here we remark that in the semiclassical approximation the electron performs
a complicated trajectory inside the wire. All the semiclassical trajectories
are tangent to the caustic surfaces of a set of concentric cylinders inside
the wire. \cite{kell} Quantization of the electronic states leads to selection
of only a certain subset of trajectories associated with a ceratin set of radii
$r_m$ of the caustic surfaces, corresponding to allowed values of the
azimuthal quantum numbers, $m$, i.e, $\kappa r_m = m$; this description
is closely related to the semiclassical periodic orbit theory. \cite{brac}
In the course of developing semiclassical methods, Keller and Rubinow
\cite{kell} have demonstrated that the Debye asymptotic expansion \cite{jahn}
of the Bessel functions ($1 \ll m < \kappa r$) provides an accurate 
approximation to the eigenfunction $J_m (\kappa r)$, i.e.,
\begin{eqnarray}
J_m (r) && \sim
 ( \frac{2}{\pi} )^{1/2} ( \kappa^2 r^2 - m^2)^{-1/4} \nonumber \\
 \times && \sin \left[ (\kappa^2 r^2 - m^2)^{1/2} - 
m \arccos \left( \frac{m}{\kappa r} \right) + \frac{\pi}{4} \right]~.
\label{bes2}
\end{eqnarray}
This approximation is valid in the region between the caustic cylindrical 
surface and the boundary surface of the wire; in the region inside the caustic
surface ($m > \kappa r$) the solution decays exponentially. In this
approximation, the equation for the asymptotic values of the Bessel-function
zeroes has the form,
\begin{equation}
(\gamma_{nm}^2 - m^2)^{1/2} - m \arccos \left( \frac{m}{\gamma_{nm}} \right)
= \pi \left( n - \frac{1}{4} \right)~.
\label{zero}
\end{equation}

First we calculate the density of states whose evaluation involves,
after integration over $p_\perp$, double sums over the quantum numbers 
$n$ and $m$; $n=1,\;2,$ ..., $m=0,\;\pm1,\; \pm2,$ ... 
[see Eq.\ (\ref{elden2})]. Applying sequentially 
the Poisson summation formula to both sums and separating the oscillatory
terms (note that in our semiclassical approximation $\kappa a \gg 1$) in
complete analogy with Refs.\ \onlinecite{ding,boga}, we obtain for the
density of states (per unit length),
\begin{eqnarray}
&& D_l^{\text{osc}} (\epsilon) =  \frac{2}{\pi a \epsilon_a} \nonumber \\
&& \times \!
\sum_{M=2}^{\infty} \sum_{Q=1}^{M/2}
\frac{1}{M} \sin \!\! \left( \frac{\pi Q}{M} \right) 
\cos \!\! \left[ 2 M K a \sin \!\! \left( \frac{\pi Q}{M} \right ) 
+ \frac{\pi M}{2} \right] \nonumber \\
&& + \frac{2 \sqrt{2} } {\pi a \epsilon_a^{3/4} \epsilon^{1/4} }
\sum_{M=1}^{\infty} \frac{1}{M^{1/2}}
\sin \left[ 2 \pi M K a \! + \! \frac{\pi}{4} \right] ~,
\label{sdosc}
\end{eqnarray}
where $\epsilon_a = \hbar^2/(2m_e a^2)$, and $K$ 
is the electron wave vector. 
The two terms in Eq.\ (\ref{sdosc}) correspond to the
contribution from the point where the phase is stationary and from the
end-points in the sum (integral) over $m$ (see discussion in 
Ref.\ \onlinecite{boga}). While the second oscillatory term in Eq.\
(\ref{sdosc}) has a smaller amplitude than the first one [by a factor
of $(K a)^{1/2}$], it corresponds to an important class of
electronic states, with $m \approx K a$, localized near the surface
of the wire (the so-called whispering gallery states \cite{boga}).

Until now we discussed a uniform wire with a constant cross-sectional
radius. In a wire with a variable shape, the cross-sectional radii
depend on $z$, as discussed in connection with Eq.\ (\ref{az}).
Substituting the $z$-dependence of the radii in Eq.\ (\ref{sdosc}), i.e.,
replacing $a$ by $a(z)$, we need to perform an integration over $z$ 
[see Eq.\ (\ref{eldt})]. This integration involves evaluation of
integrals of the form,
\begin{equation}
I = \Re \int_{-L/2}^{L/2} g(z) e^{i \alpha K a(z)} dz~,
\label{inte}
\end{equation}
where for the first term in Eq.\ (\ref{sdosc}) $g(z)=a(z)$ and
$\alpha = 2 M \sin (\pi Q /M)$, and for the second one
$g(z)=\sqrt{a(z)}$ and $\alpha=2 \pi M$. The fast oscillatory character
of the exponential factor [i.e., $K a(z) \gg 1$ for all $z$] relative to
the slow variation of $g(z)$ allows us to use the standard stationary
phase method, \cite{erde} obtaining
\begin{eqnarray}
I \approx &&
\left[ \frac{2 \pi} { \alpha K a^{\prime\prime} (0) } \right]^{1/2}
g(0) \; \Re \{ \exp[ i \alpha K a(0) + i \pi/4 ] \} \nonumber \\
&& + \frac{2} { \alpha K a^\prime (L/2) } \; g(L/2) \;
\Re \left\{ -i \exp[i \alpha K a(L/2) ] \right\}~,
\label{inte2}
\end{eqnarray}
where $z=0$ is the stationary (extremum) point, the second term is the
contribution from the end-points of the integral, and primes denote
differentiation with respect to $z$. Using the above, and after simple
algebraic manipulations, we obtain for the oscillatory part of the
density of states,
\begin{eqnarray}
&&D^{\text{osc}} (\epsilon) \nonumber \\
&& = \frac{2}{\pi} \sum_{M=2}^{\infty} \sum_{Q=1}^{M/2}
\left \{ \frac{1}{ M^{3/2} }  
\left[ \sin \left( \frac{\pi Q}{M} \right) \right]^{1/2}
\left[ \frac{\pi} { K a^{\prime\prime} (0) } \right]^{1/2} 
\right. \nonumber \\
&& \times \frac{2 m_e a(0)}{\hbar^2}
\cos \left[ 2M K a(0) \sin \left( \frac{\pi Q} {M} \right) 
+ \frac{\pi}{2} \left( M+\frac{1}{2} \right) \right] \nonumber \\
&& \left. + \frac{1}{M^2} \frac{2 m_e a(L/2)} {\hbar^2 K a^\prime (L/2)}
\sin \!\! \left[ 2 M K a(L/2) \sin \!\! \left( \frac{\pi Q}{M} \right)
+ \frac{\pi M}{2} \right] \right\} \nonumber \\
&& + \frac{2 \sqrt{2}}{\pi \epsilon^{1/4}}
\sum_{M=1}^{\infty} \left\{ \frac{1}{M} 
\left[ \frac{1}{K a^{\prime\prime} (0) } \right]^{1/2}
\frac{1}{a(0)} \right. \nonumber \\ 
&& \times \left( \frac{2 m_e a^2(0)}{\hbar^2} \right)^{3/4}
\cos [ 2 \pi M K a(0) ] \nonumber \\
&& - \frac{1}{\pi M^{3/2}} \frac{1}{K a^\prime (L/2) a(L/2)} 
\left( \frac{2 m_e a^2 (L/2)}{\hbar^2} \right)^{3/4} 
\nonumber \\
&& \times \left. \cos \left[ 2 \pi M K a(L/2) 
+ \frac{\pi}{4} \right] \right\}.
\label{sdosc2}
\end{eqnarray}
The density of states of the wire contains oscillatory contributions
from the narrowmost cross-section of the wire [first and third term in
Eq.\ (\ref{sdosc2})] and from the wire's-end cross-sections (second and
fourth terms). The amplitudes of the latter oscillations is smaller.

Having obtained the expression for the oscillatory part of the 
density of states, we can calculate
now the semiclassical approximation to the grand-canonical thermodynamic
potential $\Omega$ [see Appendix A; at zero temperature,
$\Omega = \int (\epsilon - \epsilon_F) D(\epsilon) d\epsilon$]. Restricting 
ourselves for brevity to the largest contribution [that is, to the first 
term in Eq.\ (\ref{sdosc2}) corresponding to the main contribution from
the narrowmost part of the wire], we get for the oscillatory  part of 
$\Omega$,
\begin{eqnarray}
&& \Omega^{\text{osc}} \approx \nonumber \\
&& \frac{2 \epsilon_F } { \sqrt{\pi k_F a^{\prime\prime}(0)} \; a(0) }
\sum_{M=2}^{\infty} \sum_{Q=1}^{M/2} \frac{1}{M^{7/2}}
\sin^{-3/2} (\pi Q/M) 
\nonumber \\
&& \times \cos \!\! \left[ 2 M k_F a(0) \sin (\pi Q/M) 
+ \frac{\pi}{2} \left( M + \frac{1}{2} \right) \right]~.
\label{omosc}
\end{eqnarray}

From this expression, the oscillating part of the force as a function of the
length of the constricted region [i.e., $a(0)$ in general depends on
$L$, see e.g., Eq.\ (\ref{a0})] is given by 
\begin{equation}
F^{\text{osc}} (L) =
- \frac{\partial \Omega^{\text{osc}}}{\partial a(0)}
\frac{\partial a(0)}{\partial L}~,
\label{fdsem}
\end{equation}
which upon substitution of (\ref{omosc}) yields,
\begin{eqnarray}
&& F^{\text{osc}} (L) \approx  \nonumber \\
&& \frac{4 \epsilon_F k_F^{1/2} [\partial a(0)/ \partial L]}
{ \sqrt{\pi a^{\prime\prime} (0) } \; a(0)  } 
\sum_{M=2}^{\infty} \sum_{Q=1}^{M/2}
\frac{1}{M^{5/2}} \sin^{-1/2} (\pi Q/M) \nonumber \\
&& \times \cos \!\! \left[2 M k_F a(0) \sin (\pi Q/M) +
\frac{\pi}{2} \left( M - \frac{1}{2} \right) \right]~.
\label{fsem}
\end{eqnarray}

The expression for the conductance of the wire following the Landauer
formula involves evaluation of the number of transverse states in the 
narrowmost part of the wire. Following Ref.\ \onlinecite{boga2},
\begin{eqnarray}
&& G \approx \left( \frac{2 e^2}{h} \right) \!\! 
\frac{ [k_F a(0)]^2} {4} \left \{ 1 - \frac{2}{k_F a(0)}
+ \frac{8}{\sqrt{\pi}} 
\frac{1}{ [k_F a(0)]^{3/2} } \right. \nonumber \\
&& \times  \sum_{M=2}^{\infty} \sum_{Q=1}^{M/2}
\frac{1}{M^{3/2}} \sin^{1/2} (\pi Q/M) \nonumber \\
&& \times 
\left. \cos \!\! \left[ 2M k_F a(0) \sin(\pi Q/M)  + 
\frac{\pi}{2} \left( M - \frac{1}{2} \right) \right] \right\}~,
\label{gsem}
\end{eqnarray}
which can be expressed as a function of the length of the
constriction [see e.g., Eq.\ (\ref{a0}); we remark here that our
semiclassical treatment is valid for any adiabatic wire shape].

The non-oscillating contribution (coming from the first two terms in the curly
brackets) describes the Sharvin \cite{shar} conductance of the
constriction and the Weyl \cite{weyl} semiclassical corrections, 
and the third term describes conductance quantum
oscillations as a function of $a(0)$. From a comparison of the
expression for the oscillatory contribution to the force [Eq.\
(\ref{fsem})] with the oscillatory contribution to the conductance 
[Eq.\ (\ref{gsem})], the direct correlation between the two is immediately
evident, and both depend on the spectrum of transverse modes (conducting
channels) at the narrowmost part of the wire. This is in agreement with
the results shown in Fig.\ 5 obtained through the LDA-SCM method.

\section{Conclusions and Discussion}

In this paper, we extended our investigations \cite{yann5} of 
energetics, conductance, and mesoscopic forces in a jellium 
modelled nanowire (sodium) using the local-density-functional-based shell 
correction method to variable-shaped wires, i.e., containing a constricted
region modeled here by a parabolic dependence of the cross-sectional
radii in the constriction on $z$ (see Fig.\ 1). 
The results shown above, particularly, the oscillations in the total energy
of the wire as a function of the length of the variable-shaped constricted
region (and correspondingly its narrowmost width), the consequent
oscillations in the elongation force, 
the corresponding discrete sequence of magic wire configurations, and the
direct correlation between these oscillations and the stepwise quantized 
conductance of the nanowires, originate from quantization of the
electronic states (i.e., formation of subbands) due to the reduced lateral
(transverse) dimension of the nanowires. 
These results are in correspondence with our earlier LDA-SCM investigation
of jellium-modeled uniform nanowires. \cite{yann5} Moreover, in the
current study of a wire with a variable (adiabatic) shaped constriction,
we found that the oscillatory behavior of the energetic and transport
properties are governed by the subband quantization spectrum
(termed here electronic shells) at the narrowmost part of the constriction.
This characteristic is supported and corroborated by our semiclassical
analysis (section III).

We reiterate here that such oscillatory behavior,
as well as the appearance of ``magic numbers'' and ``magic configurations'' of
enhanced stability, are a general characteristic of
finite-size fermionic systems and are in direct analogy
with those found in simple-metal clusters
(as well as in $^3$He clusters \cite{yann4} and atomic nuclei\cite{stru,bm}), 
where electronic shell effects on the
energetics \cite{heer,yann1,yann2,yann3} 
(and most recently shape dynamics \cite{yann6} 
of jellium modelled clusters driven by forces obtained from shell-corrected 
energetics) have been studied for over a decade. 

While these calculations provide a useful and instructive framework, 
we remark that they are not a substitute for theories where the atomistic
nature and specific atomic arrangements are included 
\cite{land1,land2,land3,land4,barn} in evaluation of the
energetics (and dynamics) of these systems (see in particular Refs. 
\onlinecite{land4,barn}, where 
first-principles molecular-dynamics simulations of electronic spectra,
geometrical structure, atomic dynamics, electronic transport and
fluctuations in sodium 
nanowires have been discussed). 

Indeed, the atomistic structural
characteristics of nanowires \cite{land1,land2,land3}
(including the occurrence of cluster-derived 
structures of particular geometries\cite{land4,barn}), 
which may be observed through the use of
high resolution microscopy, \cite{kizu}
influence the electronic spectrum and 
transport characteristics, as well as the energetics of nanowires and
their mechanical properties and response mechanisms. In particular,
the mechanical response of materials involves structural changes
through displacement and discrete rearrangenent of the atoms. The 
mechanisms, pathways, and rates of such structural transformations
are dependent on the arrangements and coordinations of atoms, the
magnitude of structural transformation barriers, and the local shape of the
wire, as well as possible dependency on the history of the material
and the conditions of the experiment (i.e., fast versus slow
extensions). Further evidence for the discrete atomistic nature of the 
structural transformations is provided by the shape of the force variations
(compare the calculated Fig.\ 3(b) in Ref.\ \onlinecite{land1} and
Fig.\ 3 in Ref.\ \onlinecite{land2} with the measurements shown in Figs.\ 1 
and 2 in Ref.\ \onlinecite{rubi}), and the interlayer spacing period of the 
force oscillations when the wire narrows. 
While such issues are not addressed by models
which do not include the atomistic nature of the material,
the mesoscopic (in a sense universal) phenomena described
by our model are of interest, and may guide further research in the area of 
finite-size systems in the nanoscale regime.
Such further investigations include the occurrence of magic configurations
(i.e., sequences of enhanced stability specified by number of particles,
size, thickness or shape) in clusters, dots, wires, and thin films of
normal, as well as superconducting metals, and the effect of magnetic
fields which can influence the energetics in such systems
(e.g., leading to magnetostriction effects) through variations of the subband
spectra, in analogy with magnetotransport phenomena in nanowires \cite{boga0}.

Several directions for improving the model (while remaining within
a jellium framework) are possible. These include: (i) consideration of
more complex shapes. For example, in our current model the elongation
is distributed over the entire constriction throughout the process, while a
more realistic description should include a gradual concentration of the
elongation, and consequent shape variation, to the narrower part of the 
constriction as found through molecular-dynamics simulations; 
\cite{land1,land3} (ii)
use of a stabilized-jellium description \cite{perd} of the energetics
of the nanowire in order to give it certain elements of mechanical
stability. In this context, note also that from the total energy shown
in Fig.\ 4(c), and the corresponding total force [Fig.\ 5(c)], it is
evident that in our current model, except for the region of large
elongation close to the breaking point (i.e., $\Delta L/L_0
\geq 2.5$), the wire is unstable against spontaneous collapse
(that is shortening), i.e., there are
no energetic barriers against such process, while both experiments
\cite{rubi} and MD simulations \cite{land2} show that compression
of such wires requires the application of an external force.
Improvements of the model in these directions are most desirable in light
of the aforementioned experimental \cite{rubi} and MD-simulations
\cite{land2} observations that the total oscillating forces for elongation
and compression of nanowires are of opposite signs (i.e., negative and
positive, respectively), while our current (equilibrium model) is limited
to certain aspects of the tensile part of an elongation-compression
cycle; (iii) inclusion of bias voltage effects in calculations of the 
energetics and conductance of nanowires. \cite{lang,lang2} While
such effects may be expected to have little influence (particularly on the
energetics) at small
voltages, they could be of significance at larger ones.
Work in these directions is in progress in our laboratory.

\acknowledgments
This research was supported by a grant from the U.S. Department of Energy
(Grant No. FG05-86ER45234) and the AFOSR. Useful comments by W.D. Luedtke
are gratefully acknowledged. Calculations were performed at the 
Georgia Institute of Technology Center for Computational Materials Science.

\appendix
\section*{A}

In this Appendix, we discuss briefly a semiclassical treatment of 
temperature effects on the oscillatory behavior of the force and conductance 
in nanowires. The grand-canonical thermodynamic potential at finite 
temperature, $T$, is given by
\begin{equation}
\Omega = - k_B T \sum_i \ln \left[ 1 + \exp \left( 
 \frac{\mu-\epsilon_i}{k_B T} \right) \right]~,
\label{omeg}
\end{equation}
where $i$ denotes $(n,m,p_{\perp})$, and $\mu$ is the chemical potential.

From Eq.\ (\ref{omeg}), the finite temperature expressions for
$\Omega^{\text{osc}}$, $F^{\text{osc}}$, and $G^{\text{osc}}$ differ from
those given for the zero-temperature limit in Eqs.\ (\ref{omosc}),
(\ref{fsem}), and (\ref{gsem}), respectively, by a multiplicative
factor in the sums of these equations. This factor is given by \cite{note}
\begin{mathletters}
\label{tdep}
\begin{equation}
\Psi (X_{MQ}) = \frac{X_{MQ}}{\sinh ( X_{MQ} )}~,
\label{psi}
\end{equation}
where 
\begin{equation}
X_{MQ} = \frac{2 \pi M k_B T a(0) \sin( \pi Q / M) } {\hbar v_F}~,
\label{xmq}
\end{equation}
\end{mathletters}
with $v_F$ being the Fermi velocity. For $T=0$, $\Psi(x)=1$. 

Note that the temperature dependence given in Eq.\ (\ref{tdep}) is valid
for systems with $k_F a(0) \gg 1$, and leads to reduction of the
oscillation amplitudes when $2 \pi M k_B T \geq \Delta \epsilon$, where
$\Delta \epsilon = \hbar v_F /[a(0) \sin (\pi Q/M)]$ is an
effective energy-level spacing of the electrons contributing to the
oscillatory parts of the thermodynamic potential, force, and conductance.

\begin{figure}
\caption{Schematic drawing of the jellium-background of a variable-shaped
nanowire. The (cylindrical) symmetry axis
of the wire is along the $z$-axis, with a constricted region
$(-L/2 \leq z \leq L/2)$ described by a dependence of the cross-sectional
radii $a(z)$ on $z$ [see Eq.\ (\protect\ref{az})]. 
$R_0=a(z=\pm L/2)$ is the radius in the uniform part of the wire outside
the constriction.
}
\end{figure}

\begin{figure}
\caption{Potential profiles at the narrowmost 
($z=0$; left curve) and end-points ($z=\pm L/2$; right curve)
of a constriction with an elongation $\Delta L /L_0 = 1.125$ (whose 
${\cal O}$-parameters are $L_0=80$ a.u. and $R_0=25$ a.u.). 
The local transverse-mode spectra, $\widetilde{\epsilon}_{nm}$, 
associated with these two profiles are also displayed along the left and right
$y$-axes. The dashed line indicates the Fermi level.
For the ($z=0$)-profile, the spectrum is labeled with the
corresponding $n,m$ local transverse eigenvalues ($n$ denotes the number 
of radial nodes plus one, and $m$ denotes the azimuthal angular-momentum
quantum number). The same spectrum is numbered sequentially  
(in parentheses) to
facilitate comparison between the location of the energy levels  
and the numbered peaks in the density of states given in the lower curve
in Fig.\ 3(a). Energies in units of eV, and lengths in a.u.
}
\end{figure}
			
\begin{figure}
\caption{Densities of states for: (a) two configurations of the 
variable-shaped wire, one (lower curve) with elongation $\Delta L /L_0=1.125$ 
and narrowmost radius $a_0=12.62$ a.u. (potential profiles 
and local transverse spectra for this case
are displayed in Fig.\ 2), and the other with elongation $\Delta L
/L_0 = 2.75$ and narrowmost radius $a_0=4.57$ a.u. 
(upper curve whose $y$-axis is shown on the right).
(b) a uniform-in-shape wire with $\Delta L/L_0 =0$ and $a_0=R_0=25$ a.u. 
For all cases, $L_0=80$ a.u. and $R_0=25$ a.u.
The vertical dashed lines denote the corresponding Fermi levels.
The Fermi level of the constrictions, which for the uniform-in-shape wire is 
$-2.82$ eV, varies only by 0.05 eV for all the elongations down to the
breakage point. In (a) the numbered peaks correspond to the locations of
the transverse energy levels in the narrowmost part of the constriction
[e.g., compare the lower curve in (a) with the spectrum shown on the left
axis of Fig.\ 2]. The arrows indicate the locations of some of the
transverse energy levels at the end-points of the constriction,
coinciding with corresponding peaks in the spectrum of the uniform-in-shape
wire shown in (b).
}
\end{figure}

\begin{figure}
\caption{Energies (in eV units) of a variable-shaped sodium nanowire,
plotted versus the relative elongation $\Delta L /L_0$. The initial
parameters ${\cal O}$ are $L_0=80$ a.u. and $R_0=25$ a.u. The 
smooth, ETF contribution $(\widetilde{E})$, the shell correction
($\Delta E_{\text{sh}}$), and the total energy ($E_T$) are displayed
in (a), (b), and (c), respectively.
}
\end{figure}

\begin{figure}
\caption{(a-c): The smooth, ETF contribution to the force ($\widetilde{F}$),
shell-correction force ($\Delta F_{\text{sh}}$), and total force
($F_T$), corresponding to the energies shown in Fig.\ 4(a-c), plotted
versus the relative elongation $\Delta L / L_0$. The arrow in (c) indicates
the point $F_T=0$ corresponding to the local minimum in the total
energy shown in Fig.\ 4(c). The dashed lines indicate the zeroes of the
$y$-axes. Forces in units of nanonewtons.
(d): The conductance, $G$, for the variable-shaped wire
in units of $g_0=2e^2/h$, plotted versus $\Delta L / L_0$, 
evaluated as described by Eq.\ (\protect\ref{land}).
(e): The variation of the cross-sectional radius (in a.u.) of the
narrowmost part of the constriction, plotted versus
$\Delta L / L_0$ [see Eq.\ (\protect\ref{a0}), 
with $L_0=80$ a.u. and $R_0=25$ a.u.]
}
\end{figure}

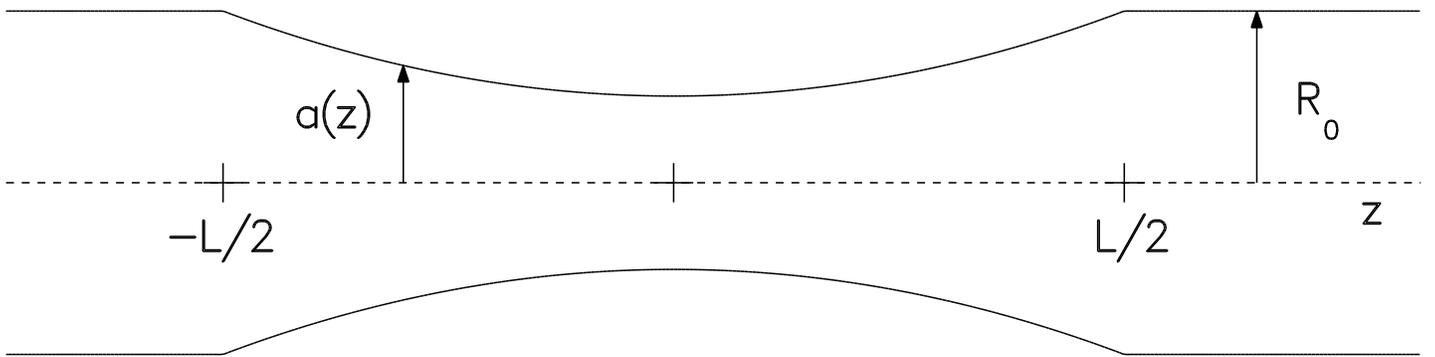
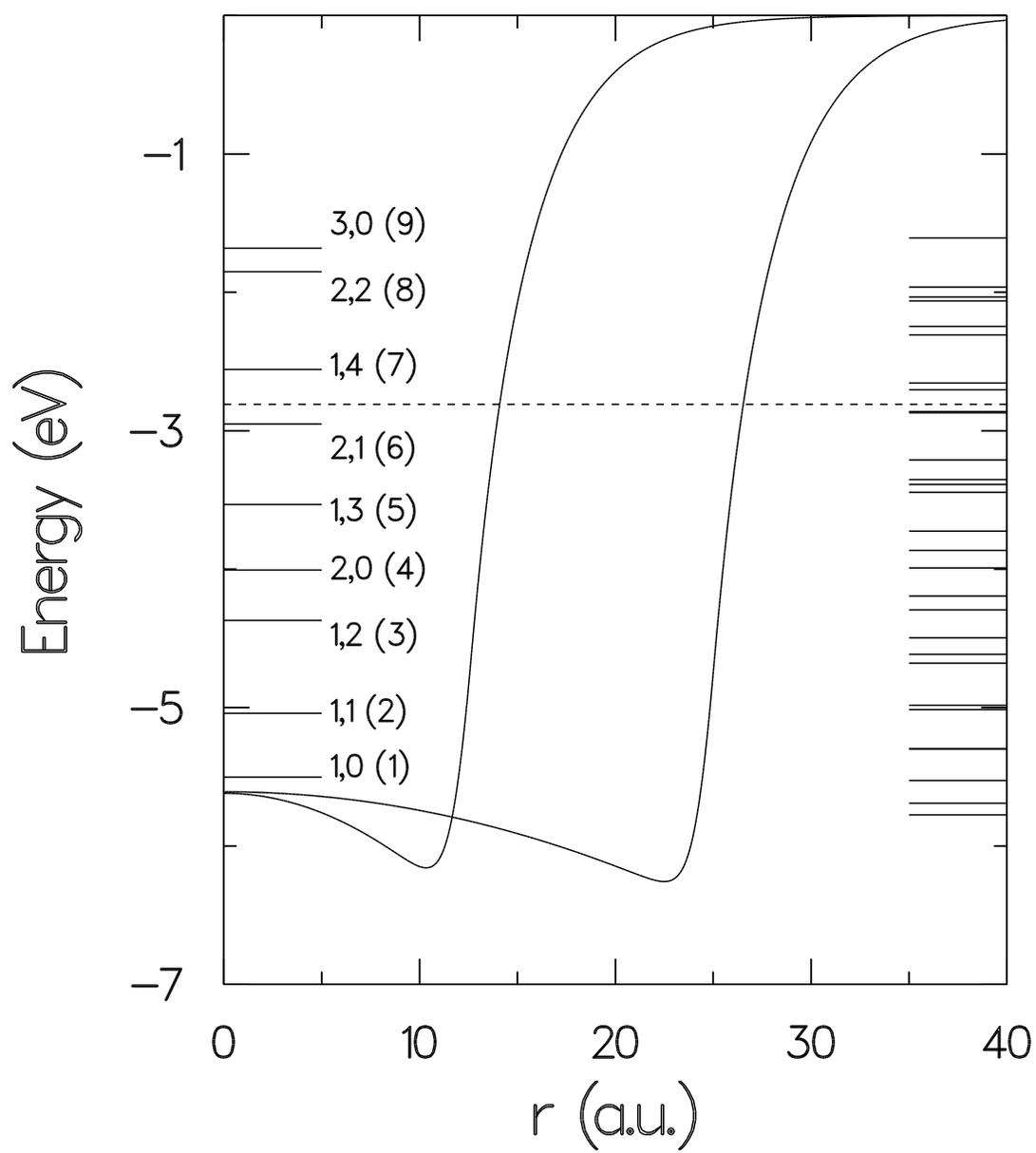
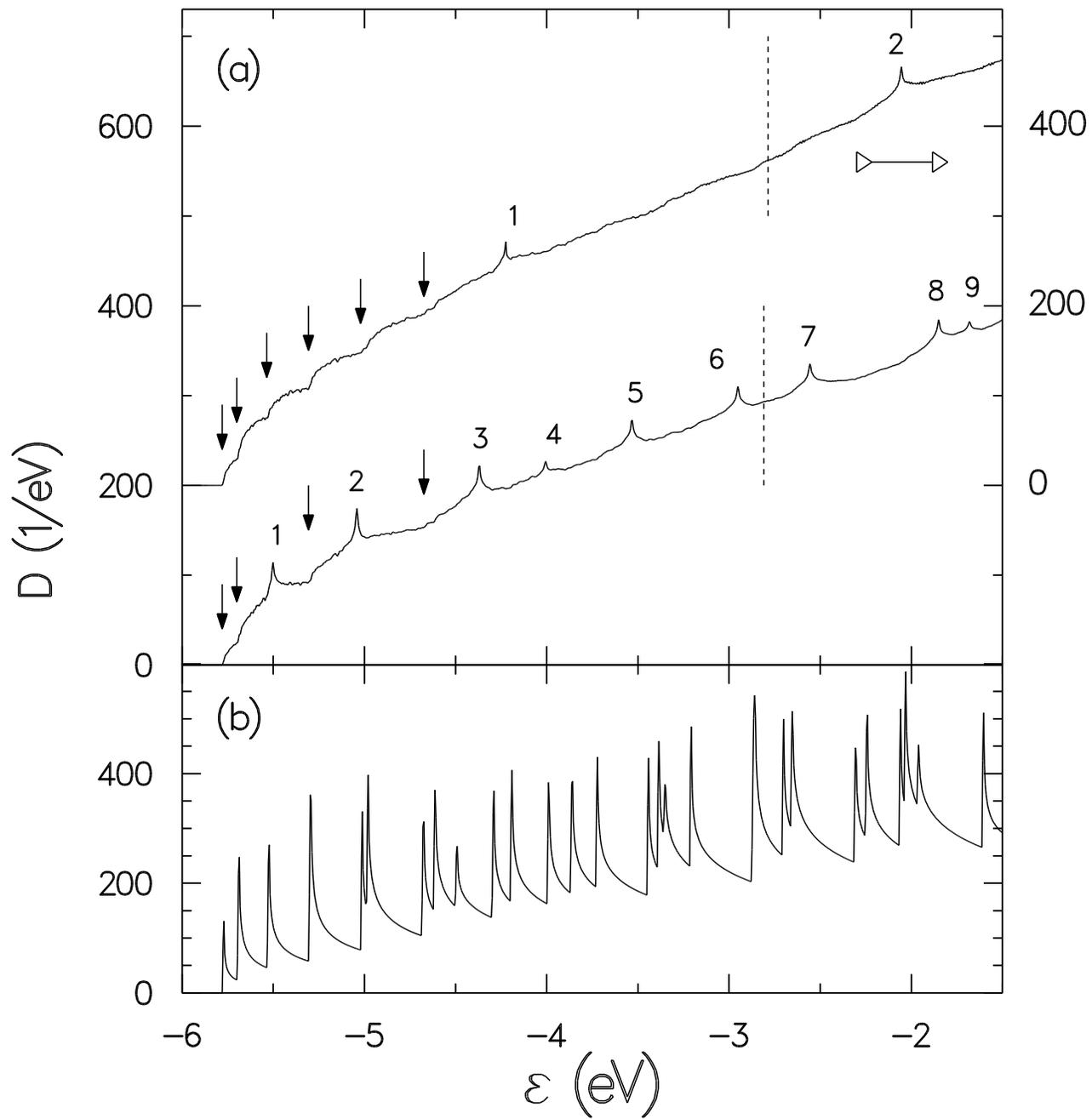
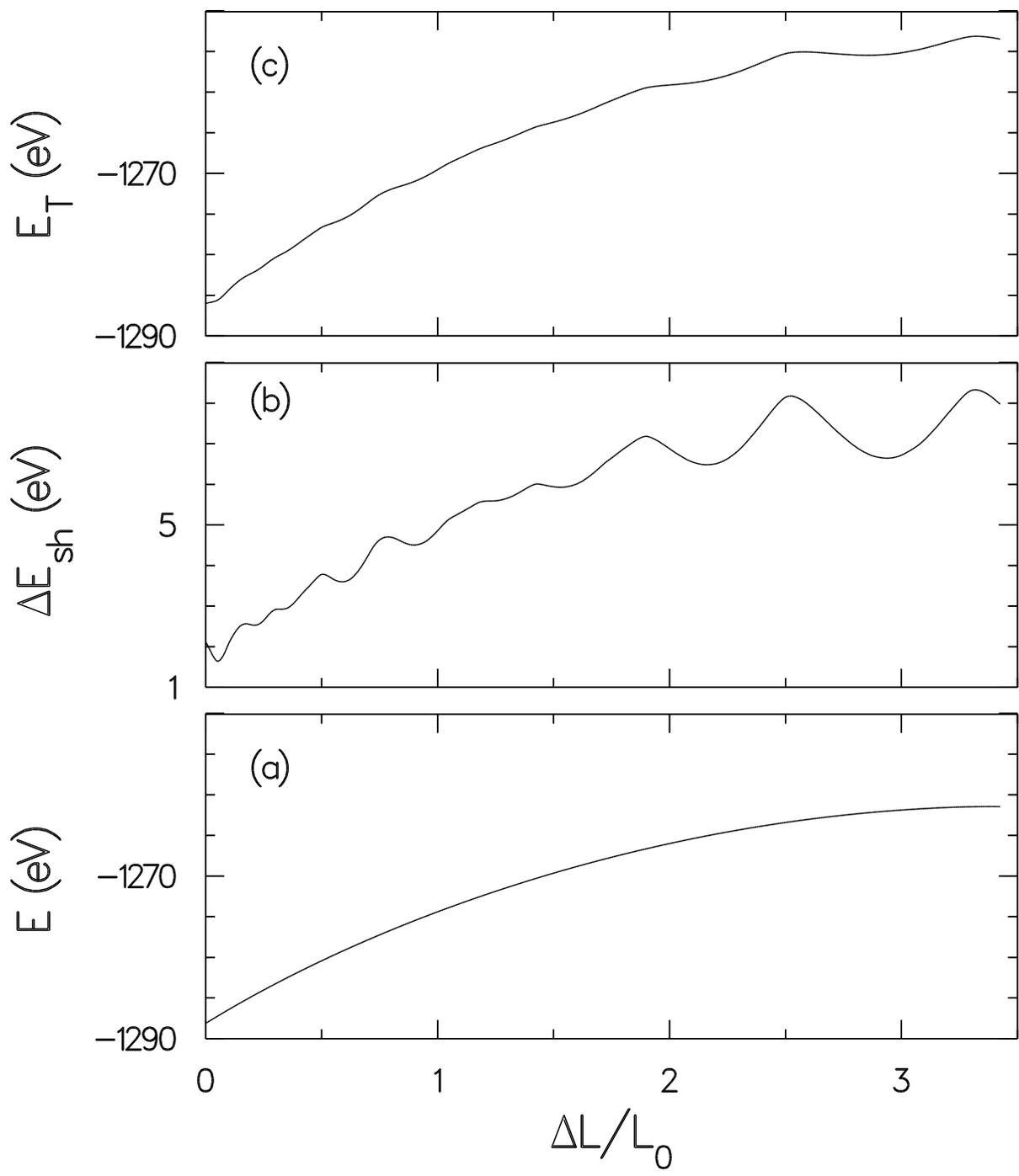
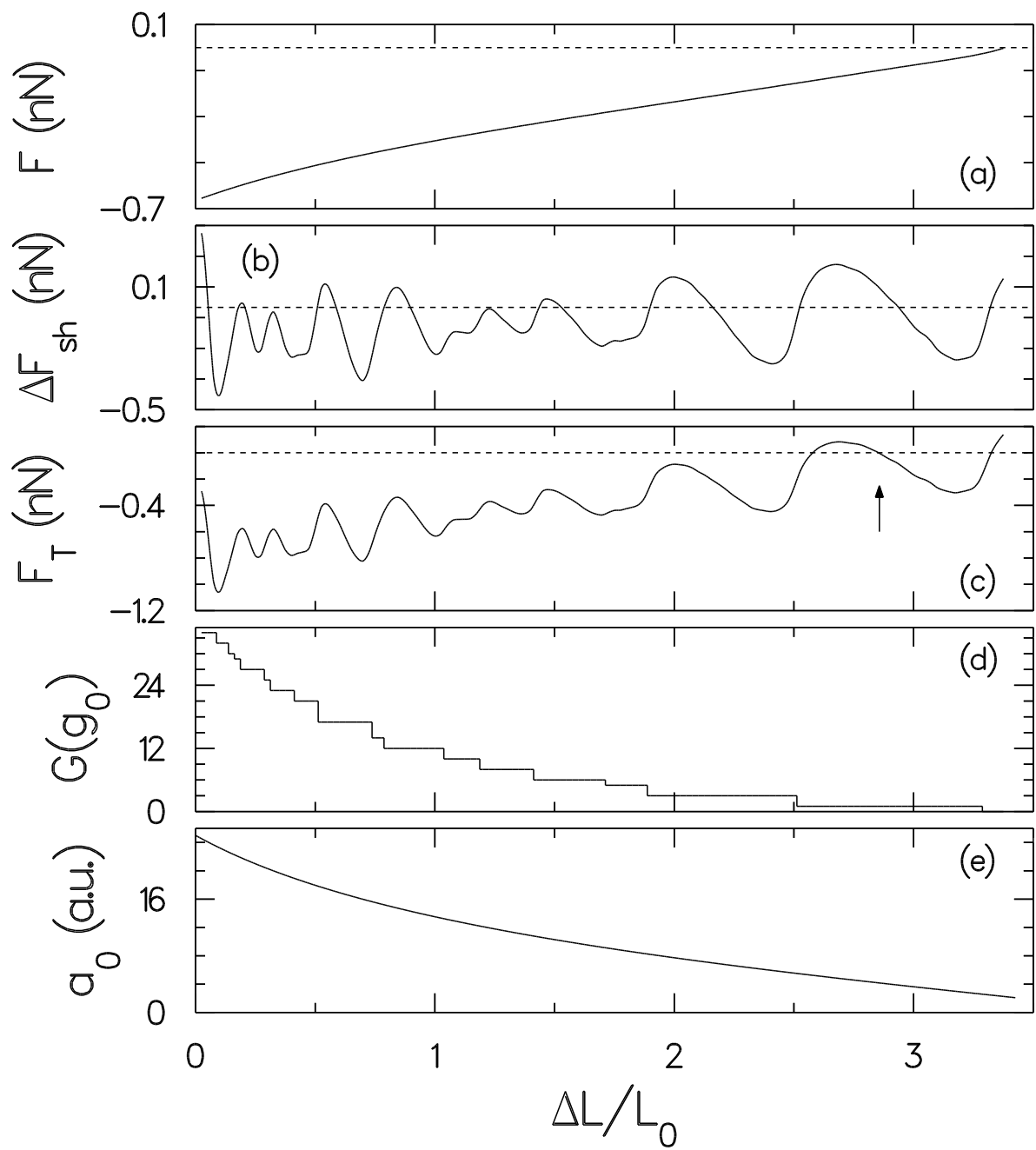
\end{document}